\documentclass[cameraready]{Interspeech}
\usepackage{tabularx}
\usepackage{makecell} 
\usepackage{placeins}
\newcolumntype{Y}{>{\centering\arraybackslash}X}

\title{WQ-Fusion: Dynamic Gated Attention for Cross-Domain Audio Representation}

\author[affiliation={1}, orcid=0009-0007-8691-2113]{Mingda}{Lin}

\author[affiliation={1},orcid=0009-0003-8259-207X]{Lei}{Ding}
\author[affiliation={1}, orcid=0009-0002-8219-7048]{Xinyue}{Zhou}
\author[affiliation={1}, orcid=0009-0001-7873-1191]{Tiantian}{Xiong}
\author[affiliation={1}, orcid=0009-0006-6372-1464]{Hanchen}{Pei}

\author[affiliation={1}, orcid=0000-0002-6825-7473, correspondingauthor]{\protect\\Gongping}{Huang}
\author[affiliation={2}, orcid=0000-0002-6749-8320]{Hao}{Zhang}
\author[affiliation={3}, orcid=0000-0003-0083-9247]{Jingdong}{Chen}
\author[affiliation={4}, orcid=0000-0002-0036-5865]{Jacob}{Benesty}

\address{
    $^1$ School of Electronic Information, Wuhan University, Wuhan, Hubei, China \\
    $^2$ Tencent AI Lab Seattle, Seattle, USA \\
    $^3$ CIAIC, Northwestern Polytechnical University, Xi’an, Shaanxi, China\\
    $^4$ INRS-EMT, University of Quebec, Montreal, Canada
}

\email{mingda.lin@whu.edu.cn,  gongpinghuang@whu.edu.cn}

\keywords{speech recognition, human-computer interaction, computational paralinguistics}

\usepackage{comment}

\begin{document}

\maketitle

\begin{abstract}
    While pre-trained models excel in specialized tasks, learning universal representations across diverse acoustic domains remains challenging. To address this, we propose WQ-Fusion, a robust dual-encoder framework for cross-domain audio representation learning. Overcoming the limitations of static concatenation, WQ-Fusion integrates whisper and qwen via an Adaptive Feature Modulation module and a novel element-wise gated attention mechanism. This design enables dynamic feature selection, allowing the model to selectively emphasize relevant acoustic and semantic dimensions. Extensive experiments on the Interspeech 2026 Audio Encoder Capability Challenge (Track A) benchmark demonstrate that by effectively routing heterogeneous information, WQ-Fusion achieves a superior overall score of 0.836, significantly outperforming the strongest single-encoder baseline.
\end{abstract}

\section{Introduction}

As a fundamental component of audio signal processing systems, audio encoders are widely used across diverse downstream applications, including speech processing, sound analysis, and music information retrieval (MIR)~\cite{kong2020panns}. From a perceptual perspective, the quality of the learned audio representations largely determines the upper bound of performance in downstream tasks~\cite{liu2022audio}. For example, audio encoders extract discriminative acoustic features to support tasks such as keyword spotting, speaker attribute analysis, and audio authenticity detection. In more complex scenarios that require higher-level understanding and reasoning, they provide essential representations for tasks including acoustic scene recognition and automatic music tagging.

With the rise of generative artificial intelligence, the role of audio encoders is undergoing a paradigm shift from end-to-end perception toward cross-modal decoupling \cite{radford2021learning}. However, this shift also poses a stringent challenge to the universality of audio representations: Audio signals exhibit pronounced heterogeneity in both spectral characteristics and temporal structure; consequently, representations optimized for speech phonetic structure often struggle to capture the rich harmonic timbre of music or the transient, impulsive patterns of acoustic events. This intrinsic tension between ``task specificity'' and ``representational universality'' has become a central bottleneck in building truly all-scenario, general-purpose audio encoders.

Existing paradigms diverge in their architectural designs and optimization goals, shaping distinct representational strengths. Speech-centric self-supervised models, such as WavLM~\cite{chen2022wavlm} and HuBERT~\cite{hsu2021hubert}, utilize masked prediction and denoising to capture phonetic structures and complex acoustic relations. Alternatively, generative acoustic models like AudioMAE~\cite{huang2022masked} leverage high-ratio masked autoencoding to reconstruct spectrogram patches, capturing a broad spectrum of non-speech textures. Weakly supervised models, exemplified by Whisper-large~\cite{radford2023robust}, achieve robust speech-language alignment through large-scale training on massive audio-text pairs. Recently, language-augmented foundation models such as Qwen2-Audio~\cite{chu2024qwen2} and MiDashengLM~\cite{dinkel2025midashenglm} have integrated audio encoders with Large Language Models (LLMs) to achieve comprehensive interactive understanding. Despite these advances, relying on a single encoder inevitably introduces specific inductive biases, failing to simultaneously reconcile high-fidelity acoustic details with deep semantic abstraction.

Building on these observations, the Interspeech 2026 Audio Encoder Capability Challenge (Track A)\footnote{\url{https://dataoceanai.github.io/Interspeech2026-Audio-Encoder-Challenge/}} provides a standardized benchmark for evaluating general-purpose audio representations. In unified generative evaluation framework, our systematic benchmarking of several representative encoder families reveals a strong representational complementarity between different paradigms. Remarkably, even a naive concatenation of Whisper-large~\cite{radford2023robust} (Whisper) and Qwen2-Audio-7B~\cite{chu2024qwen2} (Qwen) already outperforms the strongest single-encoder baselines.

However, such static fusion strategies are inherently rigid, uniformly processing all inputs without dynamically adapting to the specific demands of heterogeneous downstream tasks. To fully unlock the synergy between these encoders, we propose WQ-Fusion, a robust dual-encoder framework for cross-domain audio representation learning. Going beyond simple concatenation, WQ-Fusion introduces an Adaptive Feature Modulation module and a novel element-wise gated attention mechanism. This design endows the model with the capacity for dynamic feature selection, allowing it to adaptively emphasize or suppress specific acoustic and semantic dimensions based on contextual relevance. Experimental results demonstrate that by effectively routing cross-model information, WQ-Fusion achieves a superior overall score of 0.836, establishing a highly effective paradigm for universal audio representation.

\section{Related Work}

\subsection{Foundational Audio Encoders and Universal Representations}

Recent progress in audio representation learning has seen a convergence toward unified single-encoder architectures, each characterized by distinct inductive biases. Frameworks like AudioMAE~\cite{huang2022masked} prioritize reconstruction-based acoustic modeling to capture fine-grained textures, while Whisper~\cite{radford2023robust} and Qwen~\cite{chu2024qwen2} leverage large-scale weak supervision and LLM-integration to achieve robust semantic alignment. Parallel to these, neural codecs such as AUV~\cite{chen2025auv}, UniCodec~\cite{jiang2025unicodec} and Sematicodecc~\cite{liu2024semanticodec} attempt to unify diverse audio domains within a single framework through specialized vector-quantized (VQ) codebook structures. Together, these methodologies represent a significant leap toward general-purpose audio understanding.
However, these paradigms each face inherent representational gaps that hinder a truly holistic audio understanding. Acoustic-centric models often struggle with high-level semantic reasoning, whereas speech-centric or LLM-augmented models may underrepresent non-speech nuances or lose direct acoustic grounding. VQ-based approaches attempt to bridge this divide by integrating semantic and acoustic information within a shared discrete space, yet the discretization process inherently introduces precision loss, which prevents the model from achieving fine-grained perception of complex audio signals.

\subsection{Fusion of Heterogeneous Encoders}

Given the complementary strengths of the aforementioned models, effectively fusing heterogeneous backbone encoders has emerged as a compelling route toward universal audio representation. However, prevailing strategies for such integration often necessitate intensive large-scale fine-tuning and intricate heuristic engineering to reconcile disparate representation spaces~\cite{bharadwaj2026cmu}. Although empirically successful, these supervision-heavy approaches often obscure the intrinsic contributions of the underlying representational capacities of the encoders. Furthermore, such fusion pipelines are frequently designed as static or rigid structures, which struggle to adaptively prioritize relevant features across diverse and fluctuating acoustic contexts. This lack of architectural flexibility motivates the exploration of more dynamic integration strategies that can autonomously arbitrate between divergent inductive biases.

\subsection{Gated Mechanism}
To overcome the limitations of static integration, dynamic routing and gated attention mechanisms have recently garnered significant attention across the machine learning landscape. In the realm of Large Language Models (LLMs), recent breakthroughs have demonstrated that gated attention can offer superior non-linearity and sparsity while effectively mitigating the attention-sink problem, thereby significantly enhancing training stability and model performance~\cite{qiu2025gated}. This success extends to multimodal and audio-visual domains, where gated cross-attention has proven highly effective in dynamically adjusting modality weights based on contextual relevance. For instance, dynamic cross-attention and router-gated fusion have been successfully applied to audio-visual person verification~\cite{praveen2024dynamic} and noise-robust speech recognition~\cite{lim2025improving} to suppress irrelevant interference. Similarly, gated attention architectures have been widely adopted to enhance multimodal alignment in complex reasoning tasks, such as multimodal emotion recognition~\cite{he2025gia}, offensive content detection~\cite{hossain2025co}, financial movement prediction~\cite{zong2025stock}, and medical diagnosis~\cite{ortiz2025cognialign}. Inspired by these dynamic modulation paradigms, we hypothesize that the fusion of heterogeneous audio encoders should be a context-aware selection process rather than a fixed projection, allowing the model to adaptively route phonetic and semantic information.

\section{Method}

\begin{figure}[!t]
  \centering
  \includegraphics[width=0.46\textwidth]{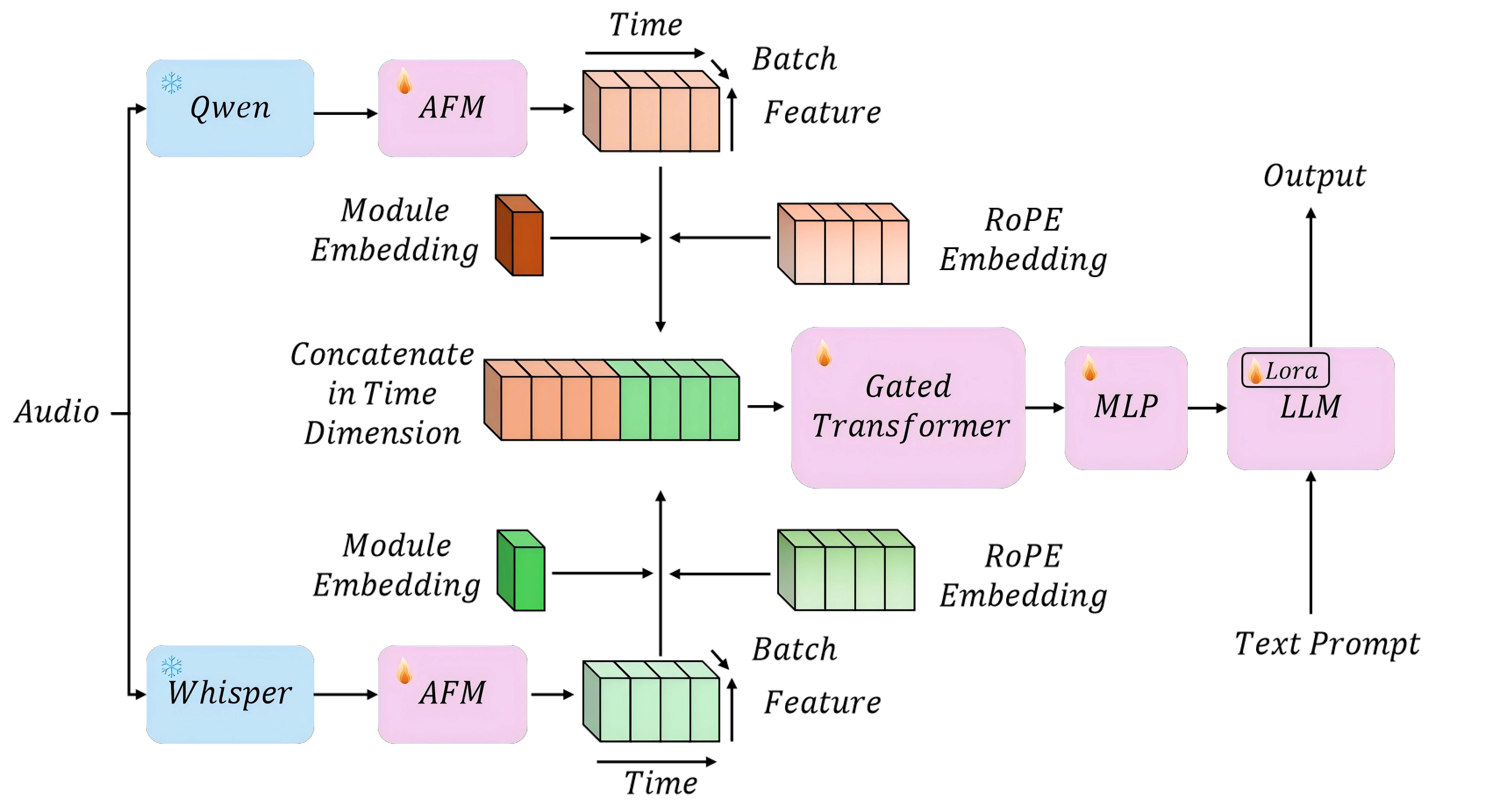}
  \caption{Architecture of WQ-Fusion.}
  \label{fig:framework1}
\end{figure}

This section details the architectural design of WQ-Fusion, a dual-encoder framework engineered to bridge the gap between phonetic-centric and semantic-centric audio representations. As illustrated in Figure 1, the pipeline operates through four strategic stages: (i) the selection of complementary backbone encoders; (ii) feature harmonization via an Adaptive Feature Modulation (AFM) module; (iii) the injection of position encoding; and (iv) dynamic information routing using a Gated Transformer mechanism.

\subsection{Backbone Encoder Selection}
To establish a robust foundation for universal audio representation, we conducted an extensive empirical evaluation of contemporary open-source encoders, encompassing self-supervised models (\textit{e.g.}, AudioMAE), acoustic event experts (\textit{e.g.}, ATST-F), music-optimized encoders (\textit{e.g.}, MuQ, MERT), and large-scale speech-language models (\textit{e.g.}, Whisper, Qwen). Our benchmarking on the XARES-LLM protocol reveals that Whisper and Qwen exhibit the most superior performance, characterized by a distinct and highly complementary relationship.

Specifically, the Whisper encoder excels in capturing fine-grained phonetic structures and linguistic nuances, a strength rooted in its extensive weak supervision on speech-centric datasets. In contrast, Qwen benefits from multi-stage optimization across diverse corpora, demonstrating exceptional proficiency in non-speech audio domains and high-level semantic reasoning. By synergizing these two divergent inductive biases, WQ-Fusion is designed to preserve high-fidelity acoustic details while maintaining deep semantic coherence across heterogeneous tasks.

\subsection{Adaptive Feature Modulation} To effectively harmonize the representations extracted from the Whisper and Qwen backbones, we developed an adaptation module based on the Adaptive Feature Modulation (AFM) mechanism. Drawing inspiration from feature modulation techniques like FiLM~\cite{perez2018film}, which are widely utilized in multimodal learning, our module avoids rigid projections by predicting dynamic scale and shift parameters for the incoming features. For an input feature sequence $X^{(\mathrm{i})}$ from encoder $i \in \{\text{Whisper, Qwen}\}$, the system first applies normalization,  with a linear projection network to predict the scale parameter $\gamma^{(\mathrm{i})}$ and the shift parameter $\beta^{(\mathrm{i})}$ afterwards. The final adapted feature $X_{\mathrm{adapt}}^{(\mathrm{i})}$ is then computed as:
\begin{align}& \hat{X}^{(\mathrm{i})} = Norm(X^{(\mathrm{i})}) \label{eq:norm}, \\
& [\gamma^{(\mathrm{i})}, \beta^{(\mathrm{i})}] = Linear(X^{(\mathrm{i})}), \label{eq:linear} \\
& X_{\mathrm{adapt}}^{(\mathrm{i})} = \gamma^{(\mathrm{i})} \odot \hat{X}^{(\mathrm{i})} + \beta^{(\mathrm{i})}. \label{eq:adapt}\end{align}
where $\odot$ denotes element-wise multiplication. Following this process, the modulated features from both the Whisper and Qwen encoders are forwarded to the downstream network for further integration.

\subsection{Positional Encoding Methods} 

To endow the model with the capability of recognizing token positions and origins, we incorporate a hybrid encoding scheme consisting of Rotary Position Embedding (RoPE)~\cite{su2024roformer} and learnable module embeddings.

Regarding temporal encoding, since the output sequences of both Whisper and Qwen possess identical lengths, tokens at the same index naturally correspond to the same acoustic segment. We thus independently apply RoPE to each encoder feature sequence to inject positional information by rotating feature vectors in 2D subspaces. Unlike traditional absolute encodings, this approach effectively preserves relative distance and temporal dependencies during self-attention.

Furthermore, to enable the model to differentiate between inputs from heterogeneous backbones, we introduce distinct learnable module embeddings $E_{\mathrm{i}}$ which are analogous to the segment embeddings used in BERT~\cite{devlin2019bert}. These embeddings are broadcasted across the temporal dimension and added to the RoPE-transformed adapted features. 
Given the input $X_{\mathrm{adapt}}^{\mathrm{(i)}}$ from encoder $i \in \{\text{Whisper, Qwen}\}$, the positionally-enriched representation is formulated as:
\begin{align}X_{\mathrm{emb}}^{\mathrm{(i)}} = RoPE(X_{\mathrm{adapt}}^{\mathrm{(i)}}) + E_{\mathrm{i}}. \label{eq:pos_encoding}\end{align}

The final integrated feature sequence $X_{\mathrm{emb}}$ is then constructed by concatenating these two representations along the temporal dimension. This design ensures that the subsequent Gated Transformer can dynamically perceive both the temporal order and the architectural source of the incoming signals, facilitating informed arbitration between phonetic and semantic information.

\subsection{Gated Transformer Mechanism} 

Based on the fused feature representations $X_{\mathrm{emb}}$, we implement an element-wise gated attention mechanism to perform dynamic feature selection across the integrated sequences, which follows the gated attention paradigm for large language models proposed by Qiu et al.~\cite{qiu2025gated}. Specifically, we augment the query projection layer to simultaneously generate both the attention queries and the gating signals. For a given input $X_{\mathrm{emb}}$, the projections for the Query $Q$, the Gate $G$, the Key $K$, and the Value $V$ are formulated as follows:
\begin{align}
& Q, G = W_{\mathrm{q}} X_{\mathrm{emb}}, \label{eq:qg_projection} \\
& K = W_{\mathrm{k}} X_{\mathrm{emb}}, \label{eq:k_projection} \\
& V = W_{\mathrm{v}} X_{\mathrm{emb}}. \label{eq:v_projection}
\end{align}
In this formulation, $W_{\mathrm{k}}$ and $W_{\mathrm{v}}$ represent the standard transformation matrices for the key and value projections, while $W_{\mathrm{q}} \in \mathbb{R}^{d \times 2d}$ serves as an augmented weight matrix that simultaneously yields a concatenated output of queries and gating tokens. To modulate the information flow, the gating signal is passed through an activation function to produce an element-wise mask. The gated attention output is then computed by:
\begin{align}
& Attention(Q, K, V) = Softmax\left(\frac{QK^T}{\sqrt{d_{\mathrm{k}}}}\right)V, \label{eq:attention} \\
& X_{\mathrm{out}} = Attention(Q, K, V) \odot \sigma(G). \label{eq:gate_fusion}
\end{align}
where $\odot$ denotes the element-wise Hadamard product. By incorporating this gating mechanism directly into the attention pipeline, the model gains the capacity for dynamic feature selection. This allows the framework to selectively emphasize or suppress specific dimensions of the combined Whisper and Qwen representations based on their contextual relevance. Such a design significantly enhances the model's ability to utilize heterogeneous information, ultimately leading to the superior performance observed in our cross-domain evaluations.

\begin{table*}[t!]
\centering
\caption{Performance comparison of single encoders and fusion strategies.}
\label{tab:ablation}
\footnotesize 
\setlength{\tabcolsep}{2pt} 
\renewcommand{\arraystretch}{0.98}
\begin{tabularx}{\textwidth}{ll*{8}{Y}} 
\toprule
 & & \multicolumn{4}{c}{Single} & \multicolumn{4}{c}{Fusion} \\
\cmidrule(lr){3-6} \cmidrule(lr){7-10} 
Domain & Task & \makecell{Dasheng\\-Base} & AudioMAE & \makecell{Whisper\\-Large} & \makecell{Qwen2\\-Audio-7B} & \makecell{Concat.} & \makecell{Adapt. and\\Trans.} & \makecell{Gated\\Trans.} & \makecell{WQ-Fusion} \\
\midrule
Speech & SC~\cite{warden2018speech}           & 0.655 & 0.472 & 0.746 & 0.792 & 0.788 & \textbf{0.955} & \underline{0.952} & 0.938 \\
Speech & LibriCount~\cite{stoter2018libricount}                & 0.386 & 0.476 & 0.469 & 0.508 & 0.503 & \underline{0.579} & \underline{0.579} & \textbf{0.583} \\
Speech & VoxLingua107~\cite{valk2021voxlingua107}              & 0.311 & 0.144 & 0.970 & 0.883 & 0.970 & 0.971 & \textbf{0.980} & \underline{0.975} \\
Speech & VoxCeleb1~\cite{nagrani2020voxceleb}                 & 0.974 & 0.595 & 0.958 & 0.969 & 0.982 & \underline{0.983} & 0.982 & \textbf{0.985} \\
Speech & ASVspoof~\cite{kinnunen2018automatic}                & 0.937 & 0.916 & 0.986 & \textbf{0.991} & 0.982 & 0.962 & 0.973 & \underline{0.979} \\
Speech & FSC~\cite{lugosch2019speech}    & 0.984 & 0.782 & 0.941 & \underline{0.994} & \underline{0.994} & 0.993 & 0.989 & \textbf{0.995} \\
Speech & VocalSound~\cite{gong2022vocalsound}                & 0.855 & 0.909 & 0.916 & 0.930 & \underline{0.945} & 0.938 & 0.936 & \textbf{0.938} \\
Speech & CREMA-D~\cite{cao2014crema}                         & 0.621 & 0.526 & 0.702 & 0.815 & \textbf{0.849} & \underline{0.842} & 0.825 & 0.820 \\
\midrule
Sound  & ESC-50~\cite{piczak2015esc}                         & 0.755 & 0.757 & 0.802 & 0.863 & \underline{0.917} & 0.909 & 0.906 & \textbf{0.930} \\
Sound  & FSD50k~\cite{fonseca2021fsd50k}                     & 0.063 & 0.143 & 0.173 & 0.252 & \underline{0.293} & 0.258 & 0.278 & \textbf{0.295} \\
Sound  & UrbanSound 8k~\cite{salamon2014dataset}             & 0.829 & 0.854 & 0.834 & 0.847 & 0.857 & 0.869 & \textbf{0.872} & \underline{0.871} \\
Sound  & FSD18-Kaggle~\cite{fonseca2018general}              & 0.415 & 0.682 & 0.719 & 0.766 & \textbf{0.838} & 0.787 & 0.821 & \underline{0.828} \\
\midrule
Music  & GTZAN~\cite{sturm2013gtzan}                         & 0.323 & 0.808 & 0.808 & 0.919 & 0.899 & \textbf{0.932} & \underline{0.930} & 0.929 \\
Music  & NSynth-I~\cite{engel2017neural}                     & 0.675 & \underline{0.757} & 0.698 & 0.743 & \textbf{0.768} & 0.735 & 0.746 & 0.748 \\
Music  & FMA~\cite{defferrard2016fma}                        & 0.429 & 0.605 & 0.585 & 0.660 & 0.710 & 0.716 & \underline{0.718} & \textbf{0.725} \\
\midrule
-- & Overall & 0.614 & 0.628 & 0.754 & 0.796 & 0.820 & 0.829 & \underline{0.832} & \textbf{0.836} \\
\bottomrule
\end{tabularx}
\end{table*}

\section{Experiment}


\subsection{Dataset}
To evaluate the cross-domain robustness of WQ-Fusion, we conduct experiments on a comprehensive benchmark comprising 15 datasets across three domains. The Speech Domain comprises Speech Commands (SC)~\cite{warden2018speech}, LibriCount~\cite{stoter2018libricount}, VoxLingua107~\cite{valk2021voxlingua107}, VoxCeleb1~\cite{nagrani2020voxceleb}, ASVspoof~\cite{kinnunen2018automatic}, Fluent Speech Commands (FSC)~\cite{lugosch2019speech}, VocalSound~\cite{gong2022vocalsound}, and CREMA-D~\cite{cao2014crema}. The Sound Domain includes ESC-50~\cite{piczak2015esc}, FSD50k~\cite{fonseca2021fsd50k}, UrbanSound 8k~\cite{salamon2014dataset}, and FSD18-Kaggle~\cite{fonseca2018general}. The Music Domain involves GTZAN Genre~\cite{sturm2013gtzan}, NSynth-Instruments (NSynth-I)~\cite{engel2017neural}, and Free Music Archive Small (FMA)~\cite{defferrard2016fma}. These datasets utilize audio-label pairs for mainstream acoustic classification tasks.

\subsection{Model Training} To ensure a fair and consistent comparison, our training protocol strictly adheres to the baseline configuration established by the competition organizers, optimizing the model for 100,000 steps with a batch size of 4 on a curated mixture of diverse acoustic datasets. While the backbone encoders and the core Large Language Model remain frozen to preserve their extensive pre-trained knowledge, we perform task-specific fine-tuning on a specific set of lightweight components. These trainable parameters encompass the MLP-based projection layers situated before the LLM, the Low-Rank Adaptation (LoRA)~\cite{hu2022lora} matrices integrated within the LLM layers, and our proposed components, including the Self Adaptation Modules, Learnable Module Embeddings, and the Gated Transformer fusion architecture. The fusion of Whisper and Qwen features is facilitated by a single-layer Gated Transformer block leveraging a multi-head mechanism with 8 attention heads. Its hidden dimension is fixed at 1280, ensuring architectural parity with the output sizes of both encoders.

\section{Results and Ablation}

\subsection{Main Evaluation Results}
The comprehensive evaluation results presented in Table~\ref{tab:ablation} provide a detailed comparison between single-encoder baselines and various fusion strategies across speech, sound, and music tasks.

Among the single-encoder models, Qwen demonstrates the most robust cross-domain generalization, achieving the highest overall baseline score of 0.796. This performance is complemented by speech-language pre-trained models such as Dasheng and Whisper, which exhibit specialized expertise in phonetic and linguistic tasks. Specifically, Whisper reaches a high of 0.986 on ASVspoof, while Dasheng maintains competitive results across speaker-related benchmarks despite its lower overall average of 0.614. Conversely, the more acoustic-oriented AudioMAE focuses on low-level feature reconstruction and yields an overall score of 0.628. Such performance divergence highlights the domain-specific inductive biases inherent in individual architectures, as no single network is capable of simultaneously capturing high-fidelity acoustic textures and deep semantic abstractions.

To address these individual limitations, we investigate the integration of heterogeneous encoders by directly concatenating the representations from Whisper and Qwen. This straightforward fusion results in a significant leap to an overall score of 0.820, which exceeds the best single-encoder performance by a substantial margin. These results confirm that phonetic-centric and semantic-centric encoders provide highly non-redundant and complementary information. However, this static concatenation remains fundamentally rigid as it cannot adaptively prioritize relevant features for different acoustic environments. By applying the proposed WQ-Fusion framework to introduce dynamic routing, the overall performance is further elevated to a state-of-the-art score of 0.836.

\subsection{Ablation Study on Fusion Strategies}
To explicitly quantify the impact of our architectural innovations, we conduct a step-wise ablation study focusing on the transition from static to dynamic integration.

The baseline fusion, utilizing a simple Concatenation approach, establishes a performance floor of 0.820. The introduction of an Adaptation and Original Transformer module~\cite{vaswani2017attention} improves the overall metric to 0.829. Alternatively, utilizing our Gate Transformer as the primary fusion mechanism in the absence of the adaptation module yields a higher performance of 0.832. This result underscores that our gating mechanism is inherently better suited for regulating the contributions of distinct encoders than vanilla self-attention.

Finally, the complete WQ-Fusion framework achieves the optimal score of 0.836 by systematically integrating the Adaptive Feature Modulation module with element-wise gated attention. As evidenced by the detailed task scores, WQ-Fusion not only secures the top overall metric but also demonstrates remarkable robustness in specific domains. These results solidly validate that dynamic, context-aware feature selection is essential for resolving the representational tension between diverse acoustic domains.

\FloatBarrier

\section{Conclusion}

This work introduces WQ-Fusion, a novel framework for universal audio representation that leverages the complementary strengths of encoders trained on diverse datasets and learning paradigms. Built upon frozen Whisper and Qwen backbones, the system introduces an Adaptive Feature Modulation mechanism to harmoniously align heterogeneous embeddings through dynamic scaling and shifting. Furthermore, to overcome the rigidity of traditional static integration, the framework incorporates a Gated Transformer architecture that utilizes element-wise gating to adaptively route information based on specific acoustic contexts. Experimental results across 15 diverse datasets demonstrate that WQ-Fusion significantly elevates the overall benchmark score from 0.796 to 0.836, demonstrating the effectiveness of cross-model fusion for audio representation learning. The proposed lightweight architecture enables scalable and robust universal audio encoding without resource-intensive backbone fine-tuning.

\section{Acknowledgments}
This work was supported by the National Natural Science Foundation (NSFC) of China under Grant 62471340. The numerical calculations in this paper have been done on the supercomputing system in the Supercomputing Center of Wuhan University.



\section{Use of Generative AI Disclosure}
In the preparation of this manuscript, generative AI tools were utilized exclusively for English translation and language polishing. All research concepts, experimental designs, data analyses, and the core scientific content were independently developed and produced by the human authors. The authors take full responsibility for the originality, validity, and final content of this paper.




\bibliographystyle{IEEEtran}

\bibliography{mybib}

\end{document}